\documentclass[sigplan,nonacm]{acmart}
\usepackage{mathtools}
\usepackage{booktabs}
\usepackage{multirow}
\usepackage{xurl}
\usepackage{enumitem}
\usepackage{siunitx}
\usepackage{balance}
\usepackage{placeins}
\usepackage{cleveref}
\hypersetup{hidelinks}
\usepackage[ruled,linesnumbered]{algorithm2e}
\usepackage{float}
\usepackage{setspace}
\usepackage{xcolor}
\usepackage{tabularx}
\usepackage{booktabs}
\usepackage{array}
\definecolor{darkred}{HTML}{9E1B1B}
\definecolor{cmtgray}{HTML}{5C5C5C}

\SetArgSty{textnormal}
\SetKwComment{xcom}{}{}
\SetCommentSty{textnormal}

\SetKwProg{Proc}{procedure}{}{}
\newcommand{\moedispatch}{\textbf{Dispatch}}
\newcommand{\moecombine}{\textbf{Combine}}
\graphicspath{{figures/}}

\begin{document}

\title[X-Stage]{X-Stage: Modeling Post-Issue Backpressure in GPU Communication--Computation Fusion}
\author{Jianwen Xian}
\affiliation{%
  \institution{KlingAI Research}
  \country{China}
}
\email{xianjianwen@klingai.com}

\author{Zhiyuan Xu}
\affiliation{%
  \institution{Tsinghua University}
  \country{China}
}
\email{xuzy24@mails.tsinghua.edu.cn}

\author{Yuchen Li}
\affiliation{%
  \institution{Tsinghua University}
  \country{China}
}
\email{liyuchen24@mails.tsinghua.edu.cn}

\author{Ziliang Lai}
\affiliation{%
  \institution{KlingAI Research}
  \country{China}
}
\email{laiziliang@klingai.com}

\author{Jie Li}
\affiliation{%
  \institution{KlingAI Research}
  \country{China}
}
\email{lijie@klingai.com}

\author{Kang He}
\affiliation{%
  \institution{KlingAI Research}
  \country{China}
}
\email{hekang@klingai.com}

\author{Yilin Zhang}
\affiliation{%
  \institution{NVIDIA}
  \country{United States}
}
\email{yilzhang@nvidia.com}

\author{Zhen Huang}
\affiliation{%
  \institution{NVIDIA}
  \country{United States}
}
\email{zhhuang@nvidia.com}

\author{Qinqin Chen}
\affiliation{%
  \institution{NVIDIA}
  \country{United States}
}
\email{cheric@nvidia.com}

\author{Aichen Feng}
\affiliation{%
  \institution{NVIDIA}
  \country{United States}
}
\email{aichenf@nvidia.com}

\author{Jinyan Chen}
\affiliation{%
  \institution{NVIDIA}
  \country{United States}
}
\email{jinyanc@nvidia.com}

\author{Julien Lai}
\affiliation{%
  \institution{NVIDIA}
  \country{United States}
}
\email{julienl@nvidia.com}

\author{Chengru Song}
\affiliation{%
  \institution{KlingAI Research}
  \country{China}
}
\email{songchengru@klingai.com}

\renewcommand{\shortauthors}{Xian et al.}

%

\makeatletter

\renewcommand{\@mkauthors}{%
  \global\setbox\mktitle@bx=\vbox{%
    \noindent
    \unvbox\mktitle@bx
    \par\medskip
    \centering
    {\normalsize\normalfont
    \begin{tabular}{@{}c@{}}
    Jianwen Xian$^{1}$ \quad
    Zhiyuan Xu$^{2,*,\ddagger}$ \quad
    Yuchen Li$^{2,*,\ddagger}$ \quad
    Ziliang Lai$^{1}$ \quad
    Jie Li$^{1}$ \\[2pt]
    Kang He$^{1}$ \quad
    Yilin Zhang$^{3}$ \quad
    Zhen Huang$^{3}$ \quad
    Qinqin Chen$^{3}$ \\[2pt]
    Aichen Feng$^{3}$ \quad
    Jinyan Chen$^{3}$ \quad
    Julien Lai$^{3}$ \quad
    Chengru Song$^{1,\dagger}$ \\[5pt]
    {\small
    $^{1}$KlingAI Research
    \qquad
    $^{2}$Tsinghua University
    \qquad
    $^{3}$NVIDIA}
    \end{tabular}\par}
    \medskip
  }%
}

\makeatother

\renewcommand{\shortauthors}{Xian et al.}

\makeatletter

\begin{abstract}
Fine-grained, device-initiated communication allows fused GPU kernels to issue remote stores directly from their compute pipelines, a pattern increasingly used in expert parallelism (EP), tensor parallelism (TP), and Ulysses-style sequence parallelism (UP). Existing designs reason about where communication is issued and when remote data becomes ready, but  lack a quantitative model of the sender-side interval after a remote store is accepted and before it becomes visible at the destination. This interval determines whether communication remains decoupled from computation or backpressures it. We identify \textbf{X-Stage}, a software-visible post-issue stage with finite decoupling. Downstream pressure can dissipate while the issuer resumes useful work, whereas sustained injection consumes X-Stage headroom and eventually stalls the compute pipeline. We characterize this behavior and build a calibrated model that predicts whether remote-store arrivals accumulate backpressure or recover during intervening computation. Guided by the model, we reshape bursty arrivals when they would exhaust X-Stage headroom and exploit natural compute windows when headroom can recover concurrently. 
Evaluation across representative EP, TP, and UP workloads shows up to
1.62$\times$ fused-kernel, 1.75$\times$ end-to-end, and 1.43$\times$
sender-visible speedup, respectively. Microbenchmarks further validate the model’s predictions of backlog accumulation, recovery, and sender-side backpressure.

\end{abstract}


\maketitle

\begingroup
\renewcommand{\thefootnote}{\fnsymbol{footnote}}
\setcounter{footnote}{0}

\footnotetext[1]{These authors contributed equally to this work.}
\footnotetext[2]{Corresponding author}
\footnotetext[3]{Work done during an internship at KlingAI Research.}

\endgroup

\section{Introduction}
\label{sec:introduction}

The gap between GPU compute throughput and intra-node interconnect bandwidth
continues to widen across recent accelerator generations. Communication is
therefore increasingly exposed in tensor, expert, and sequence parallelism,
especially in distributed inference with short, low-precision compute phases
\cite{flux,comet,ulysses,megascale}. To recover overlap, systems have moved from operator- and stream-level concurrency toward fine-grained, in-kernel scheduling, often within persistent kernels \cite{tilelink,megascale,parallelkittens}. Peer mappings and symmetric-memory interfaces further allow these kernels to issue one-sided remote stores directly from the device, without returning to the host \cite{nvshmem,nvshmem_demystify}.

This shift changes the unit at which overlap must be reasoned about. Operator-level models treat communication as a bulk interval and estimate overlap from aggregate compute and transfer times. A persistent kernel instead emits a sequence of remote-store bursts---batches of writes to a peer GPU---interleaved with tile-level computation. Two schedules may move the same bytes and perform the same computation yet expose different sender-side stalls, depending on how those bursts are spaced. Aggregate bandwidth and consumer readiness alone therefore do not determine the compute critical path.

Existing fine-grained fusion systems primarily specify which role issues each transfer, when it is issued, and when a remote consumer may use the data. What remains missing from their scheduling abstractions is a quantitative model of the sender-side interval after a remote store is accepted but before it becomes remotely visible. Without such a model, a scheduler cannot determine how far the issuer can advance ahead of completion, whether  downstream pressure dissipates during useful work, or when finite downstream headroom is exhausted and backpressure propagates to the compute pipeline.

This omission produces a counterintuitive pattern in persistent GEMM-like kernels. An isolated GEMM output phase (epilogue) can issue a short remote-store burst and release its temporary output buffer quickly, making communication appear decoupled from computation. When successive epilogues issue such bursts back to back, however, later issues lengthen and eventually stall the upstream Tensor Core producer. Neither a completion-coupled model, which blocks the issuer until remote visibility, nor an unbounded fire-and-forget model explains both regimes.

\begin{figure*}[t]
  \centering
  \includegraphics[width=0.96\textwidth]{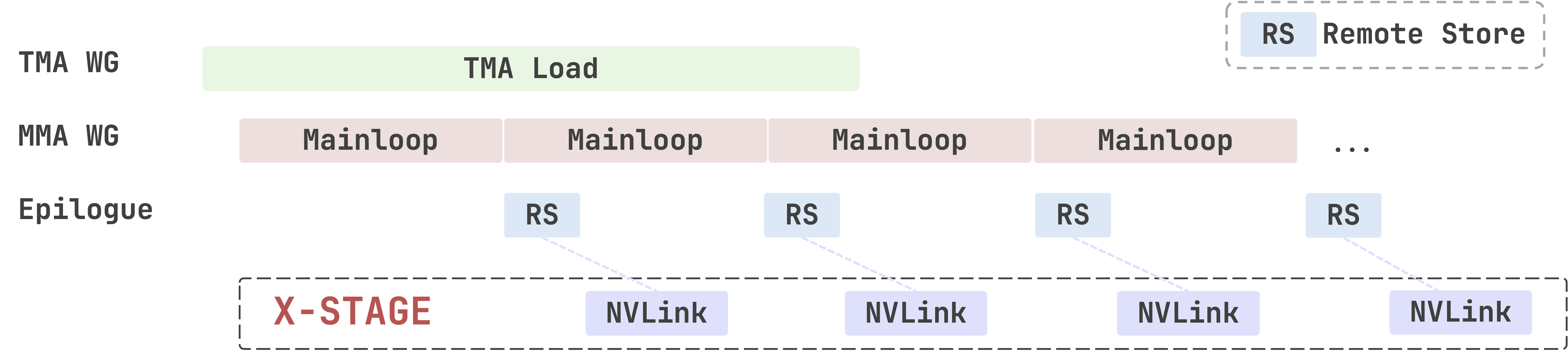}
  \caption{\textbf{X-Stage in a fused GEMM pipeline.}
  An epilogue issues a remote-store burst, after which downstream pressure can dissipate while the next mainloop performs useful work. A short gap preserves
  residual backlog; a sufficient gap restores X-Stage headroom before the next
  burst.}
  \Description{A fused GEMM timeline in which an epilogue remote-store burst
  drains through X-Stage concurrently with a subsequent compute mainloop.}
  \label{fig:overview}
\end{figure*}

We identify this missing software-visible stage as \textbf{X-Stage}. As Figure~\ref{fig:overview} illustrates, the measured downstream pressure can dissipate while the issuer resumes computation without injecting additional remote traffic. This creates an opportunity: useful computation can overlap post-issue recovery. It also imposes a constraint: the available lead is finite, so repeated bursts can consume X-Stage headroom and propagate downstream pressure back to the compute producer. X-Stage is therefore neither completion-coupled nor an unbounded fire-and-forget path. It exposes a hardware--software interface: the platform determines the finite drain and headroom, while the kernel schedule determines the arriving bursts and the computation available between them.

We characterize X-Stage with a calibrated Burst--Gap model that captures backpressure-free issue, effective downstream drain, and finite post-issue headroom. These platform parameters are calibrated independently of applications, while each kernel supplies only its remote-store arrivals and intervening compute windows. The model then predicts whether backlog accumulates across bursts, recovers during useful work, or exhausts the available headroom. It thereby turns X-Stage from a post-hoc explanation into a model-to-design audit.

The audit yields two complementary design actions. When repeated bursts consume
headroom, software reshapes arrivals with ready computation; when existing
computation restores headroom, the issuer resumes useful work during that
natural drain window. We evaluate representative EP, TP, and UP kernels.
MegaMoE achieves up to $1.62\times$ fused-kernel speedup, GEMM--AllReduce up to
$1.75\times$ end-to-end speedup, and FlashAttention--All-to-All up to
$1.43\times$ sender-visible speedup. Per-burst issue spans, producer stalls,
and residual scaling in the EP and UP cases validate the two predicted
mechanisms. A supplementary GEMM--All-to-All experiment examines another
compute--exchange boundary.

This paper makes three contributions:

\begin{enumerate}[leftmargin=1.5em]
\item \textbf{X-Stage abstraction and architectural characterization.}
We expose finite sender-side decoupling in GPU remote-store execution
between accepted issue and remote visibility, and characterize how isolated
bursts, repeated injection, and intervening computation produce recovery
or backpressure.

\item \textbf{A calibrated scheduling model for GPU programming.}
Our Burst--Gap model separates independently calibrated platform behavior
from a kernel's remote-store arrivals and compute intervals. It predicts
backlog accumulation, recovery, and capacity pressure, giving programmers
quantitative criteria for reshaping burst arrivals and exploiting existing
compute windows for communication overlap.

\item \textbf{Cross-mode evaluation and mechanism validation.}
We evaluate compute--communication fusion across representative EP, TP,
and UP workloads. Performance results quantify the gains across these
modes, while targeted measurements of issue spans, compute stalls, and
residual scaling validate the predicted recovery and backpressure
mechanisms.
\end{enumerate}

\section{Background and Motivation}
\label{sec:background}

\subsection{Remote-Store Execution Semantics}

Modern multi-GPU nodes expose peer memory through unified virtual addressing, peer mappings, and symmetric-memory interfaces \cite{cuda_peer_access,nvshmem}. These mechanisms allow a GPU kernel to write directly into another GPU's memory without returning to the host. We distinguish three software-relevant points in this process. A remote store is \emph{sender-accepted} when the local memory path accepts the request and the issuing role can advance. It becomes \emph{remotely visible} when the request reaches the destination and updates its memory. The data becomes \emph{consumer-ready} only after the required ordering, signaling, and synchronization conditions have been satisfied.

These boundaries serve different purposes. Sender acceptance determines when the producer can resume computation, whereas consumer readiness determines when the destination may safely use the data. Correct programs must still enforce fences, signals, buffer-lifetime rules, and readiness checks; early sender progress is not a remote-completion event. X-Stage concerns the finite execution interval between sender acceptance and remote visibility, rather than replacing these correctness mechanisms.

Fine-grained fusion decomposes computation and communication into tiles or roles connected through device-side synchronization \cite{flux,comet,tilelink,megascale,parallelkittens}. Existing mechanisms express application dependencies and consumer readiness well, but do not quantify how much accepted traffic remains in flight or when it will feed pressure back into the sender.

\subsection{Remote Stores across Parallelism Modes}

Device-initiated remote stores appear in several forms of distributed execution. In \emph{expert parallelism} (EP), expert outputs are returned to the ranks that own the corresponding tokens. In \emph{tensor parallelism} (TP), partial output tiles participate in push-based All-Gather, Reduce-Scatter, or All-Reduce phases. In \emph{Ulysses parallelism} (UP), attention tensors move between sequence and head partitions through All-to-All communication \cite{flux,comet,tilelink,deepgemm,ulysses}. Although these modes differ in dataflow and correctness dependencies, they can expose the same sender-side structure: batches of remote stores separated by intervals of local computation.

The parallelism mode alone does not determine X-Stage behavior. A concrete execution may accumulate outstanding traffic, recover during computation, or alternate between the two as tile shape, producer alignment, and compute spacing change. Conversely, different parallelism modes can generate similar remote-store arrival sequences.

\subsection{Why Task-Level Overlap Is Insufficient}

At operator granularity, fused latency is often approximated as
$\max(T_{\mathrm{compute}},T_{\mathrm{comm}})$. This abstraction treats communication as a bulk interval and therefore cannot distinguish two schedules that perform the same computation and transfer the same bytes but space their remote-store bursts differently. One schedule may allow accepted traffic to drain between bursts, while another may exhaust finite downstream capacity and stall the compute pipeline.

Task-level timing also cannot reveal when sender acceptance occurs, whether accepted traffic continues to progress during computation, or how much lead can be sustained before backpressure appears. These missing quantities motivate the independent X-Stage characterization in Section~\ref{sec:xstage}.

\section{Characterizing and Modeling X-Stage}
\label{sec:xstage}

The task-level abstraction in Section~\ref{sec:background} leaves the sender-side post-issue interval unresolved but does not establish its behavior. We therefore ask
four questions directly: Can the sender resume while previously accepted traffic still occupies downstream headroom? Does that headroom recover while the sender performs computation that issues no additional remote traffic? How much outstanding traffic can accumulate before later issue backpressures? Can one calibrated
model predict when subsequent issue recovers or stalls? The following
microbenchmarks answer these questions and summarize the observed behavior as
X-Stage.

\begin{figure*}[t]
  \centering
  \includegraphics[width=0.98\textwidth]{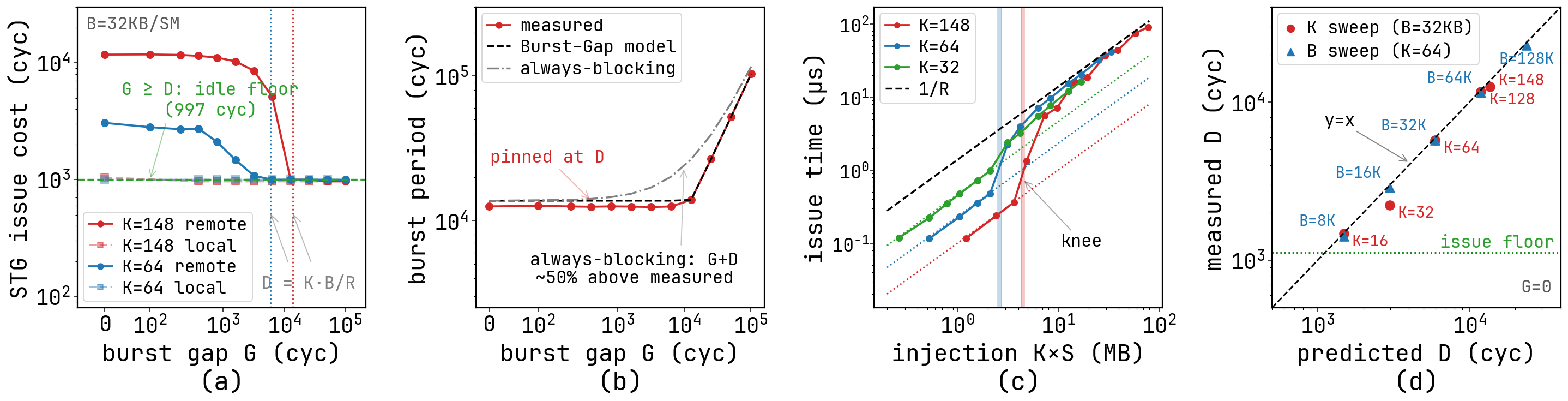}
  \caption{\textbf{X-Stage characterization.}
  (a) A producer-side non-injecting gap restores sender-visible issue time;
  (b) measured periods follow the Burst--Gap max law;
  (c) an isolated-burst knee exposes finite effective capacity; and
  (d) zero-gap periodic bursts calibrate a shared effective drain rate.}
  \Description{Four remote-store microbenchmark plots showing issue recovery,
  period prediction, isolated-burst capacity, and drain-rate calibration.}
  \label{fig:xstage-validation}
\end{figure*}

\subsection{Measurement Questions and Patterns}
\label{sec:microbench}

Our benchmark uses $\mathcal{K}$ concurrent producers. Each producer emits
$\mathcal{B}$ bytes, giving aggregate burst volume
$\mathcal{V}=\mathcal{K}\mathcal{B}$. A producer is a concurrent role issuing
remote stores; it need not be the Tensor Core producer in an application. We
measure sender-visible issue time $T_{\mathrm{iss}}$, rather than treating
remote visibility as the end of issue, and use two complementary patterns.

\textbf{Periodic bursts} alternate a remote-store burst with a producer-side
interval $\mathcal{G}$ that issues no remote stores. The interval is not a
receiver acknowledgement or a completion wait: it represents time in which an
application can execute its next mainloop while downstream pressure from previously accepted traffic can dissipate. Sweeping $\mathcal{G}$ measures recovery, while
$\mathcal{G}=0$ drives the path to a drain-limited steady state.
\textbf{Isolated bursts} start from a recovered state and sweep $\mathcal{V}$
to expose whether one burst can backpressure itself. A local-memory control
keeps the producer count, store width, address generation, and loop structure
unchanged. It separates ordinary issue-loop effects from pressure specific to
the remote path.

These patterns identify different phenomena. The periodic experiment asks
whether unfinished traffic from one burst delays the next and whether a
non-injecting interval overlaps with downstream progress. The isolated
experiment removes history and tests whether a recovered sender has finite effective outstanding headroom. Conflating the two would
make a capacity knee look like a rate limit, or make cross-period accumulation
look like a burst that is intrinsically too large.

\subsection{Observed Recovery and Drain-Limited Behavior}
\label{sec:xstage-recovery}

Figure~\ref{fig:xstage-validation} reveals three regimes. First, a recovered
short burst remains at a backpressure-free issue floor. Second, repeated bursts
with a short gap become drain-limited: increasing the gap reduces later issue
stall one-for-one until the floor is recovered. Third, a sufficiently large
isolated burst crosses a capacity knee even without previous traffic. The local
control remains nearly gap-insensitive. Together, these observations show
post-issue recovery during an interval with no additional remote-store injection, and finite effective post-issue headroom.

The shape of the period curve distinguishes X-Stage from a completion-coupled
issuer. If each burst kept its issuer blocked until remote visibility, the
period would add the gap to communication time. Instead,
Figure~\ref{fig:xstage-validation}(b) stays on a drain-limited plateau at short
gaps and changes to a recovered issue-plus-gap branch only after sufficient
recovery. Thus the gap and post-issue draining overlap; the gap is not merely
extra idle time inserted after a completed transfer. This behavioral result is
the bridge used later by the two mechanism case studies: MegaMoE inserts real Linear-1
mainloops, while FlashAttention uses the next Q-loop.

\subsection{Burst--Gap Model and Independent Calibration}
\label{sec:xstage-model}

We summarize the measured behavior with

\[
\mathcal{M}_{X}
=
\left(
T_{\mathrm{iss}}^{0}(\mathcal{K},\mathcal{V}),
\mathcal{R},
\mathcal{Q}(\mathcal{K})
\right),
\]

where $T_{\mathrm{iss}}^{0}$ is the recovered issue floor,
$\mathcal{R}$ is the effective aggregate drain rate, and $\mathcal{Q}$ is the
effective outstanding capacity. This is a calibrated fluid abstraction of
aggregate post-issue behavior, not a claim about a particular hardware FIFO.

\begin{table}[t]
  \centering
  \caption{\textbf{Burst--Gap quantities and how they are obtained.}}
  \label{tab:xstage-notation}
  \small
  \setlength{\tabcolsep}{3.5pt}
  \renewcommand{\arraystretch}{1.12}
  \begin{tabular}{@{}lp{0.68\columnwidth}@{}}
    \toprule
    \textbf{Quantity} & \textbf{Meaning / measurement} \\
    \midrule
    $\mathcal{K},\mathcal{B}$
      & Concurrent producers and bytes per producer. \\
    $\mathcal{V}$
      & Aggregate burst volume, $\mathcal{K}\mathcal{B}$. \\
    $\mathcal{G}$
      & Producer-side interval with no new remote stores. \\
    $T_{\mathrm{iss}}^{0}$
      & Recovered issue floor from the large-gap plateau. \\
    $\mathcal{R}$
      & Aggregate drain rate from zero-gap steady state. \\
    $\mathcal{Q}$
      & Effective capacity from the isolated-burst knee. \\
    $P_n,\rho_n$
      & Floor issue--gap period and normalized offered pressure. \\
    $q_n,\Delta b_n$
      & Residual backlog and its byte-level period change. \\
    $b_n^{\mathrm{iso}}$
      & Peak increment of one recovered burst. \\
    \bottomrule
  \end{tabular}
\end{table}

Assume a work-conserving effective drain at rate $\mathcal{R}$ over the
calibrated regime. In a drain-limited steady state, each period must be long
enough to drain the volume injected during that period. A period also consists
of sender-visible issue followed by $\mathcal{G}$. Combining these two views
and lower-bounding issue by its recovered floor gives the Burst--Gap relation

\begin{equation}
\begin{aligned}
T_{\mathrm{period}}
&=
\max\!\left(T_{\mathrm{iss}}^{0}+\mathcal{G},
            \frac{\mathcal{V}}{\mathcal{R}}\right),\\
T_{\mathrm{iss}}
&=
\max\!\left(T_{\mathrm{iss}}^{0},
            \frac{\mathcal{V}}{\mathcal{R}}-\mathcal{G}\right).
\end{aligned}
\label{eq:burst-gap}
\end{equation}

The minimum gap that returns issue to its recovered floor is

\begin{equation}
\mathcal{G}^{*}
=
\left[
\frac{\mathcal{V}}{\mathcal{R}}-T_{\mathrm{iss}}^{0}
\right]_{+}.
\label{eq:recovery-gap}
\end{equation}

Equation~\ref{eq:burst-gap} makes a directly testable prediction without
application-specific fitting. At short gaps,
$T_{\mathrm{period}}=\mathcal{V}/\mathcal{R}$ and every additional unit of
$\mathcal{G}$ removes approximately one unit of sender-visible issue stall. At
$\mathcal{G}^{*}$, the period switches branches; beyond it, issue remains at
$T_{\mathrm{iss}}^{0}$ and only the deliberately inserted gap grows.

We identify the parameters from disjoint regions of the experiments. The
large-gap plateau gives $T_{\mathrm{iss}}^{0}$ for a recovered burst. Zero-gap
periodic runs give $\mathcal{R}$ from
$T_{\mathrm{period}}^{\mathrm{ss}}=\mathcal{V}/\mathcal{R}$. Different
$\mathcal{K}/\mathcal{B}$ decompositions of the same aggregate volume converge
to a common drain-limited slope in
Figure~\ref{fig:xstage-validation}(d), indicating that this regime is governed
by the aggregate downstream path rather than one producer's local issue
throughput. We then use these fixed quantities to predict the complete gap
sweep in Figure~\ref{fig:xstage-validation}(a--b).

\subsection{Normalized Pressure and Backlog Recurrence}
\label{sec:xstage-recurrence}

Equation~\ref{eq:burst-gap} describes a repeated homogeneous pattern. Real
kernels produce nonuniform bursts and gaps, so we retain the same fluid
assumption but audit the sequence period by period. For burst $n$, define the
backpressure-free period and normalized offered pressure as

\begin{equation}
P_n=T_{\mathrm{iss},n}^{0}+\mathcal{G}_n,
\qquad
\rho_n=\frac{\mathcal{V}_n}{\mathcal{R}P_n}.
\label{eq:rho}
\end{equation}

$\rho_n$ is not the instantaneous issue rate inside a burst. It compares the
volume arriving over one issue--gap period with the volume X-Stage can drain
over that same period. The corresponding byte-level change is

\begin{equation}
\Delta b_n
=\mathcal{V}_n-\mathcal{R}P_n
=\mathcal{R}P_n(\rho_n-1).
\label{eq:delta-b}
\end{equation}

Let $q_n$ be the residual outstanding volume immediately before burst $n$ in
the floor-issue counterfactual. Then

\begin{equation}
q_{n+1}
=\left[q_n+\Delta b_n\right]_{+}
=\left[q_n+\mathcal{R}P_n(\rho_n-1)\right]_{+}.
\label{eq:backlog-recurrence}
\end{equation}

This relation gives $\rho_n$ and $\Delta b_n$ distinct roles: $\rho_n$ is a
dimensionless pressure that can be compared across
kernels only after calibrating the same platform and remote-store-path
configuration, whereas
$\Delta b_n$ is the actual number of bytes carried into later periods. If
every $\rho_n\le1$, no period creates positive residual backlog. This is a
simple sufficient condition, not a necessary condition for a nonuniform
schedule: an overloaded period with $\rho_n>1$ may be recovered by later
underloaded periods. Equation~\ref{eq:backlog-recurrence} captures that history
instead of assigning an entire kernel one average rate.

For a homogeneous pattern, $\rho\le1$, $\Delta b\le0$, and
$\mathcal{G}\ge\mathcal{G}^{*}$ are equivalent. The recurrence therefore
extends, rather than replaces, the directly measured Burst--Gap max law.

\subsection{Finite Effective Outstanding Capacity}
\label{sec:xstage-capacity}

An isolated burst remains backpressure-free only if its peak outstanding
volume satisfies

\begin{equation}
b_n^{\mathrm{iso}}
=
\left[\mathcal{V}_n-\mathcal{R}T_{\mathrm{iss},n}^{0}\right]_{+},
\qquad
q_n+b_n^{\mathrm{iso}}
\le \mathcal{Q}.
\label{eq:capacity-guard}
\end{equation}

The subtraction accounts for bytes that drain while the burst itself is being
issued. $b_n^{\mathrm{iso}}$ is a single-burst occupancy increment, not an
initial backlog; $q_n$ contains the history from earlier periods. If the guard
fails, actual issue must stretch until enough volume drains, so the
floor-issue recurrence has located the onset of backpressure rather than
described the already-stalled trajectory. From a recovered state $q_n=0$, the
condition reduces to the isolated-burst knee. This is why $\mathcal{Q}$ cannot
be inferred from aggregate burst size alone.

On the measured eight-GPU system, $\mathcal{R}\approx717$~GB/s. With
$\mathcal{K}=148$, Figure~\ref{fig:xstage-validation}(c) places the isolated
issue knee at about 33~KiB per producer.
Using the recovered issue floor measured
near the knee, $T_{\mathrm{iss}}^{0}\approx0.76~\si{\micro\second}$, gives an
effective capacity of $\mathcal{Q}\approx4.25$~MiB. The three parameters are therefore obtained from
the large-gap plateau, zero-gap slope, and isolated-burst knee, respectively.

\subsection{Prediction Scope}

The model captures the effective post-issue behavior of the measured
remote-store path through a recovered issue floor, an effective drain rate,
and finite outstanding capacity. For periodic burst--gap patterns,
Equation~\ref{eq:burst-gap} predicts the sender-visible issue span and
recovery knee. For nonuniform schedules, the backlog recurrence and capacity
guard evaluate the concrete burst sequence. Local staging then determines
whether the predicted issue delay reaches the compute producer.

Calibration separates these components: zero-gap runs determine the drain
rate, recovered issue measurements establish the issue floor, and isolated
bursts characterize effective capacity. Gap sweeps test the predicted
recovery knee. Within a fixed calibrated configuration, applications supply
their burst arrivals and intervening compute intervals without refitting
the X-Stage parameters.




\section{Applying X-Stage across Parallelism Modes}
\label{sec:design}

\subsection{Using X-Stage to Guide Kernel Scheduling}
\label{sec:design-audit}

A fused GEMM contains two distinct decoupling boundaries. Local accumulators
or staging buffers separate the Tensor Core producer from the epilogue
issuer; X-Stage separates accepted issue from remote-visible completion.
More local slots can delay backpressure propagation, but cannot eliminate
a sustained rate mismatch after X-Stage headroom is consumed.

We apply the same model across applications. First, calibrate
$\mathcal{M}_{X}$ for each fixed platform and remote-store-path configuration.
Next, reconstruct the kernel's burst sequence
$\{\mathcal{V}_n,T_{\mathrm{iss},n}^{0},\mathcal{G}_n\}$ and compute
$\rho_n$, $\Delta b_n$, and $q_n$ using
Equations~\ref{eq:rho}--\ref{eq:backlog-recurrence}.
Check the capacity guard
$q_n+b_n^{\mathrm{iso}}\le\mathcal{Q}$ at each burst, then determine whether
any predicted issue delay persists until the occupied local buffer must
be reused by the compute producer.

These predictions guide two scheduling choices. When repeated bursts
accumulate backlog, interleave ready, independent computation to provide
additional drain time. When existing compute windows restore headroom,
the issuer can resume useful work after issue and exploit that natural
interval for overlap. Both choices remain subject to the per-burst
capacity guard.

The resulting schedules retain the required local-buffer lifetime and
remote-consumer synchronization protocols. Headroom recovery predicts
sender-side progress; consumer readiness is established by those protocols.

\subsection{Mapping Kernels to the X-Stage Model}
\label{sec:kernel-mapping}

The mapping identifies which producers contribute to each remote-store
burst and what computation separates successive bursts.
$\mathcal{K}_{\mathrm{act}}$ denotes the number of active producers,
$\mathcal{V}_n$ their aggregate burst volume, and $\mathcal{G}_n$ the useful
compute interval before the next modeled burst. Full alignment provides
a conservative single-burst capacity check for a fixed producer set,
while the recurrence retains the actual burst order and spacing to
predict backlog accumulation and recovery.

The mapping also relates the predicted issue span to the next reuse of
the occupied local buffer. Only delay remaining at that reuse point
stalls the compute producer. In MegaMoE, interleaving changes both the
remote-store spacing and the mainloop that covers the occupied staging
slot, so both effects enter the prediction. In FlashAttention, the
completed output tile determines the burst volume, and the next Q-loop
provides the compute window. The model checks whether that burst fits
within the available capacity and whether the Q-loop restores headroom
before the next output burst.

We next apply this mapping to MegaMoE (EP),
Flash\-Attention--All-to-All (UP), and GEMM--AllReduce (TP).

\subsection{Expert Parallelism: MegaMoE in Detail}
\label{sec:megamoe-design}

MegaMoE groups local experts into waves. The original schedule generally
executes a wave's Linear-1 work and then many Linear-2 tiles. Each Linear-2
epilogue writes its result to a symmetric \moecombine{} buffer. Fine-grained
experts shorten the Linear-2 mainloop between these bursts, and aligned
epilogues create an aggregate volume
$\mathcal{V}=\mathcal{K}_{\mathrm{act}}\mathcal{V}_{\mathrm{t}}$. The
individual burst satisfies the capacity guard, but the wave's short gap gives
$\rho_n>1$ and $\Delta b_n>0$ over enough consecutive bursts for $q_n$ to
approach X-Stage capacity.

\begin{figure*}[t]
  \centering
  \includegraphics[width=0.94\textwidth]
  {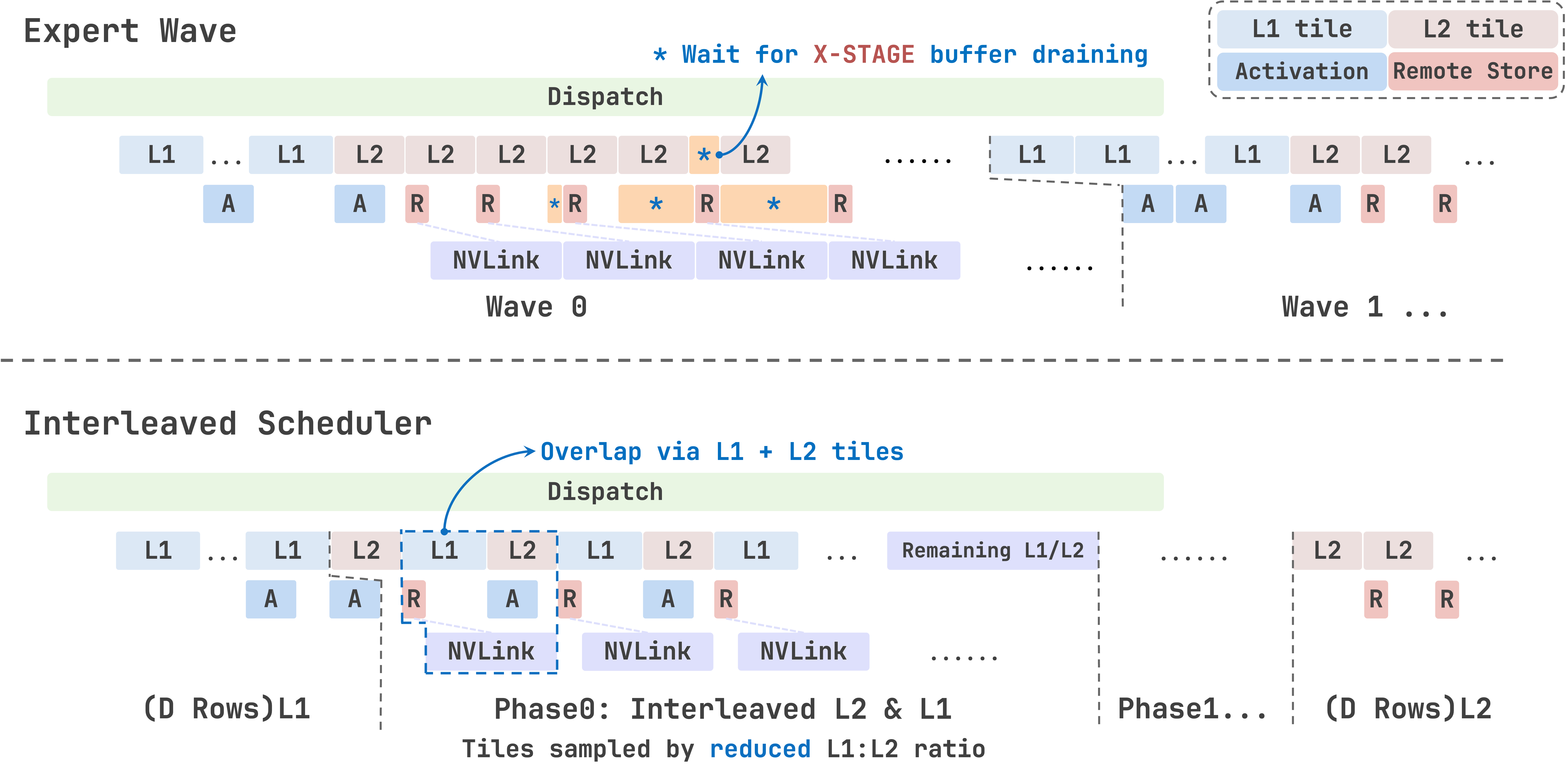}
  \caption{\textbf{Preserving X-Stage headroom in MegaMoE.}
  Expert-Wave groups Linear-2 epilogues and concentrates their remote-store
  bursts. The interleaved scheduler moves ready, independent Linear-1 tiles
  between Linear-2 tiles, creating useful drain windows without adding work or
  communication.}
  \Description{Comparison of concentrated Linear-2 remote-store bursts with an
  interleaved schedule that inserts ready Linear-1 computation.}
  \label{fig:megamoe-schedule}
\end{figure*}

Our scheduler merges the ready Linear-1 and Linear-2 tile streams across expert
waves. It emits a Linear-2 tile only when Linear-1 maintains a sufficient
dependency lead; otherwise it emits a ready Linear-1 tile. The transformation
preserves all within-expert dependencies and does not change the long-run tile
mix. If a steady cycle contains $n_1$ Linear-1 and $n_2$ Linear-2 tiles, its
average gap changes from
$\mathcal{G}_{\mathrm{wave}}\approx T_{\mathrm{L2}}^{\mathrm{mma}}$ to

\begin{equation}
\mathcal{G}_{\mathrm{int}}
\approx
T_{\mathrm{L2}}^{\mathrm{mma}}
+\frac{n_1}{n_2}T_{\mathrm{L1}}^{\mathrm{mma}}.
\label{eq:megamoe-gap}
\end{equation}

\paragraph{Trading issue stall for useful computation.}

In the drain-limited branch, inserting ready Linear-1 computation reduces
modeled issue overhead one-for-one until the recovered issue floor is
reached, without changing communication volume or the long-run tile mix.
For a homogeneous cycle satisfying the capacity guard,
$\mathcal{G}_{\mathrm{int}}\ge\mathcal{G}^{*}$ predicts recovery to that
floor. Reordering alone cannot sustain full recovery if the average
offered rate at the recovered floor still exceeds $\mathcal{R}$.
Meeting this average-rate bound is necessary but insufficient for
nonuniform schedules: their concrete burst order must also pass the
backlog recurrence and capacity guard.

\paragraph{Schedule feasibility and backpressure propagation.}

The dependency lead is determined by Linear-1 readiness across SMs,
separately from X-Stage calibration. Too small a lead makes Linear-2
wait for its inputs and disrupts interleaving. Increasing the lead beyond
the readiness requirement leaves the average-rate bound unchanged and
can enlarge the Linear-2 tail and live expert-weight working set,
harming cache locality. Eligible Linear-1 tiles must also have ready
\moedispatch{} input and a free destination ring slot.

Compute benefit depends on the delay remaining when the occupied local
staging slot must be reused. The model predicts an MMA wait equal to
the portion of the epilogue's remote-store span that exceeds the
available mainloop coverage. Interleaving can both shorten that span
and provide longer coverage through ready Linear-1 work. We therefore
measure both epilogue issue spans and MMA staging waits.
Appendix~\ref{app:megamoe} provides the supporting derivations,
scheduler pseudocode, and detailed dependency and staging analysis.


\subsection{Ulysses Parallelism: Exploit a Natural Q-Loop Gap}
\label{sec:fa-design}

Ulysses sequence parallelism converts a sequence partition to a head
partition before attention and restores the sequence partition through
a post-attention All-to-All \cite{ulysses}. We fuse this output-side
exchange with FlashAttention. Each CTA traverses all K/V tiles for its
assigned query rows before producing an output tile, placing a complete
Q-loop between successive output-side remote-store bursts.

For output tile shape $b_{\mathrm{m}}\times d$ and element size
$s_{\mathrm{out}}$, each producer emits
$\mathcal{B}_{\mathrm{t}}=b_{\mathrm{m}}d s_{\mathrm{out}}$ bytes.
Full alignment of $\mathcal{K}$ producers gives
$\mathcal{V}=\mathcal{K}\mathcal{B}_{\mathrm{t}}$ for a conservative
single-burst capacity check; the recurrence uses the actual burst order
and spacing. The intervening Q-loop supplies the useful compute gap,
which grows with sequence length.

For the evaluated FA4 configurations, the measured Q-loop exceeds the
predicted recovery gap, yielding $\rho_n<1$ and $\Delta b_n<0$.
Under the calibrated point estimates, a burst starting from recovered
headroom also satisfies the independent capacity guard, and the recurrence
predicts $q_{n+1}=0$ before the next output boundary. The existing compute
window therefore provides the modeled recovery opportunity without
requiring additional computation to reshape arrivals.

\begin{figure*}[t]
  \centering
  \includegraphics[width=0.98\textwidth]{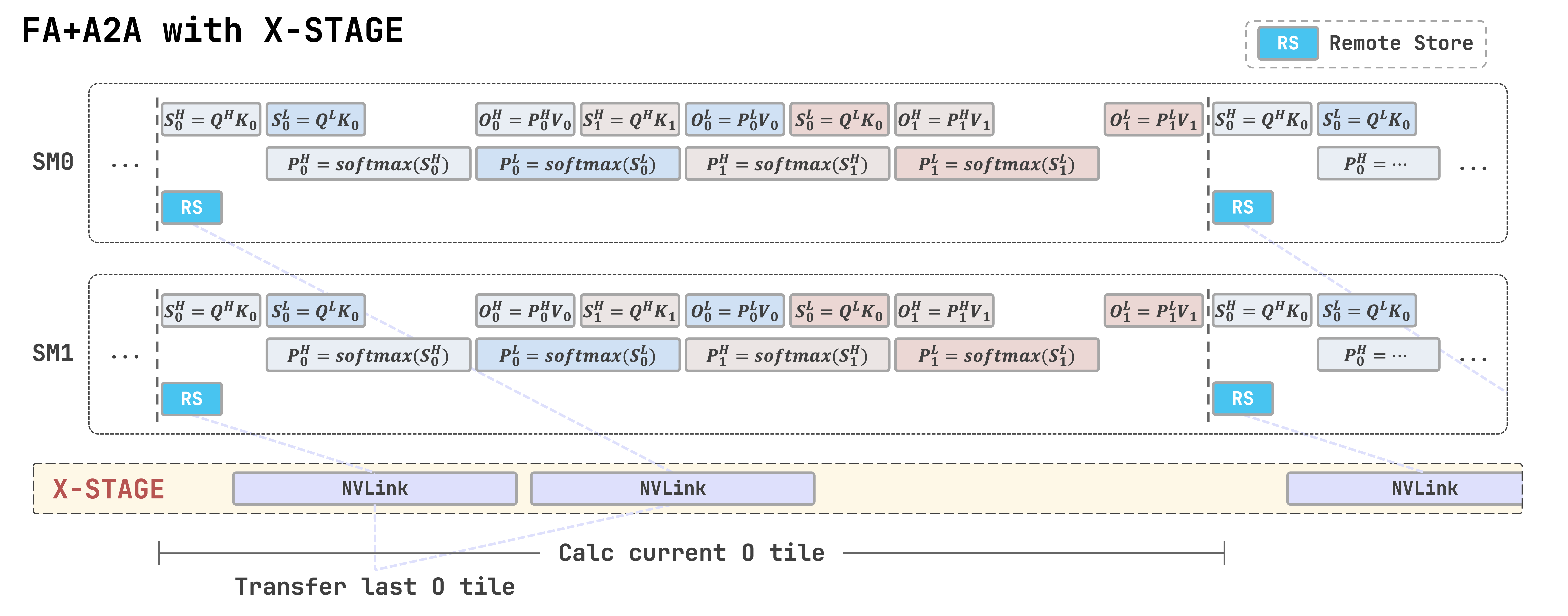}
  \caption{\textbf{Exploiting X-Stage decoupling in FlashAttention-4+A2A.}
  The output-owning role issues a tile's post-attention All-to-All stores
  and returns to computation. The next complete Q-loop provides the
  drain window, so the design reserves neither a dedicated communication
  warp nor a dedicated SM for post-issue progress.}
  \Description{A FlashAttention-4 tile pipeline in which output remote
  stores drain through X-Stage during the following Q-loop.}
  \label{fig:fa-pipeline}
\end{figure*}

This prediction guides the role assignment in
Figure~\ref{fig:fa-pipeline}. The existing output owner issues the remote
stores and returns to computation, reusing the next Q-loop for post-issue
recovery. The design reserves no dedicated communication warp or SM solely
for post-issue progress. Local-buffer lifetime and consumer-readiness
protocols remain in place; sender progress does not imply remote-visible
completion.

The resulting prediction concerns the residual latency above FA-only
execution. A completion-coupled issuer exposes a remote-completion cost
at each output boundary. X-Stage instead predicts a residual governed
by sender-visible issue work, with post-issue recovery covered by the
Q-loop. Section~\ref{sec:eval-fa} tests this distinction through
sender-visible timings and residual scaling.
Appendices~\ref{app:fa-context} and~\ref{app:fa-bound} provide the layout
illustration, parameter substitution, capacity margin, and
recovery-threshold derivation.

\subsection{Tensor Parallelism: GEMM--AllReduce Fusion}
\label{sec:tp-design}

In row-parallel tensor parallelism, a linear layer $Y=XW$ is partitioned
along its input-feature dimension. Each GPU holds a row shard $W_r$
and the corresponding activation slice $X_r$, and computes a partial
output $Y_r=X_rW_r$ through a local GEMM. For $P$ GPUs, the complete
output is $Y=\sum_{r=1}^{P}Y_r$; an AllReduce sums these contributions
and makes the result available on every rank. This GEMM--AllReduce
boundary appears in attention output projections and the second
linear layer of Transformer MLPs \cite{megatron_lm}.

We implement fusion at this boundary by initiating communication for
ready output chunks while independent GEMM work remains. The fused
execution exposes a sequence of output chunks interleaved with
computation, allowing the X-Stage model to analyze their arrival
spacing and post-issue pressure.

For a remote-store phase of the collective, aligned producers contribute
$\mathcal{V}_n=\mathcal{K}_{\mathrm{act}}\mathcal{B}_{\mathrm{c}}$,
where $\mathcal{B}_{\mathrm{c}}$ is the communicated volume per producer.
The independent computation before their next emission supplies
$\mathcal{G}_n$. AllReduce may contain several communication and reduction
phases, so the analysis follows their actual remote-store arrivals and
dependencies. Chunking reduces individual bursts but can increase their
frequency; the recurrence and capacity guard therefore apply to the
resulting sequence.

Section~\ref{sec:eval-tp} compares the fused operation with serial
GEMM+Quant followed by AllReduce. Both timings include reduction
completion, unlike the sender-visible timing scope used for
FlashAttention--All-to-All. The reported TP speedup measures the overall
fusion benefit; diagnosing its precise X-Stage regime requires the
corresponding producer and chunk trace.

\section{Evaluation}
\label{sec:evaluation}

We first test the model's recovery and capacity predictions, then evaluate
representative EP, TP, and UP workloads. MegaMoE tests whether arrival
reshaping reduces remote-store spans and upstream MMA stalls.
FlashAttention--All-to-All tests natural recovery through sender-visible
performance and residual scaling, followed by a short supplementary
GEMM--All-to-All experiment. GEMM--AllReduce then reports TP fusion performance
with collective completion included in the timing scope. These measurements use
explicit timing boundaries rather than treating all speedups as the same
metric.

\subsection{Setup and Methodology}

Experiments run on one fully connected eight-GPU NVLink/NVSwitch system with a
single NVIDIA GPU platform; each
GPU has 148 SMs. The software stack is
CUDA~\texttt{13.1}, PyTorch~\texttt{2.9}, DeepGEMM commit
\texttt{7f2a703}, and FlashAttention commit \texttt{77aacb6}. MegaMoE uses
eight-way expert parallelism. We warm up each configuration, use identical
inputs and process mappings for paired comparisons, and report medians.
Platform parameters are calibrated for the fixed platform
and remote-store-path configuration with the microbenchmarks in
Section~\ref{sec:xstage} and are not refitted to the EP or UP case studies.

CUDA events measure kernel time; Nsight Systems GPU metrics provide Tensor Core
timelines; and \texttt{clock64()} instrumentation records sender-visible
per-tile issue spans and MMA-side staging waits. MegaMoE reports the median of
20 barrier-aligned trials after taking the maximum across ranks in each trial.
For FlashAttention--A2A, the sender-side timing boundary ends after the kernel
has consumed its source staging and issued remote stores. Cross-rank barriers align measurement blocks and remain outside
the per-replay timing window during back-to-back replay.
Thus the FA
speedups quantify sender-side critical-path reduction, not early remote
readiness. Appendix~\ref{app:methodology} gives additional timing and
noise-control details.

\subsection{Model Validation and Cross-Mode Diagnosis}

The characterization itself provides the first validation. With
$\mathcal{R}$ fixed from zero-gap runs and
$T_{\mathrm{iss}}^{0}$ taken from the recovered plateau,
Figure~\ref{fig:xstage-validation}(a--b) predicts recovery and period across the
gap sweep. The isolated-burst knee in
Figure~\ref{fig:xstage-validation}(c) independently locates the finite-capacity
transition. These tests separate the three components of
$\mathcal{M}_{X}$ before applying them to kernels.

Table~\ref{tab:cross-mode-results} summarizes the evaluated fusion boundaries.
MegaMoE's expert-wave prefix accumulates modeled backlog, while the
FlashAttention Q-loop permits recovery between output bursts. For
GEMM--AllReduce, the chunk and phase mapping defines how to apply the audit;
the performance result alone does not establish its backlog trajectory.
We therefore use the EP and UP mechanism traces for the detailed model tests.

\begin{table}[t]
  \centering
  \caption{\textbf{Representative fusion cases and evaluation scope.}
  EP and UP include detailed mechanism tests; TP reports GEMM--AllReduce
  performance.}
  \label{tab:cross-mode-results}
  \small
  \setlength{\tabcolsep}{3pt}
  \renewcommand{\arraystretch}{1.12}
  \begin{tabularx}{\columnwidth}{@{}lXX@{}}
    \toprule
    \textbf{Mode} & \textbf{Fusion case} & \textbf{Evidence} \\
    \midrule
    EP & MegaMoE Combine & Arrival reshaping; kernel time. \\
    UP & FlashAttention--A2A & Natural recovery; sender time. \\
    TP & GEMM--AllReduce & Operator end-to-end time. \\
    \bottomrule
  \end{tabularx}
\end{table}

\subsection{EP/MegaMoE: Prediction, Mechanism, and Performance}
\label{sec:eval-megamoe}

We compare Expert-Wave with the interleaved scheduler across seven model shapes,
W4A8 and W8A8 precision, balanced and skewed routing, and multiple sequence and
expert configurations, for 84 configurations in total. Both versions share
compute kernels, warp specialization, buffers, epilogues, and synchronization;
only the order of ready Linear-1 and Linear-2 tiles changes.

\begin{figure*}[t]
  \centering
  \includegraphics[width=0.96\textwidth]{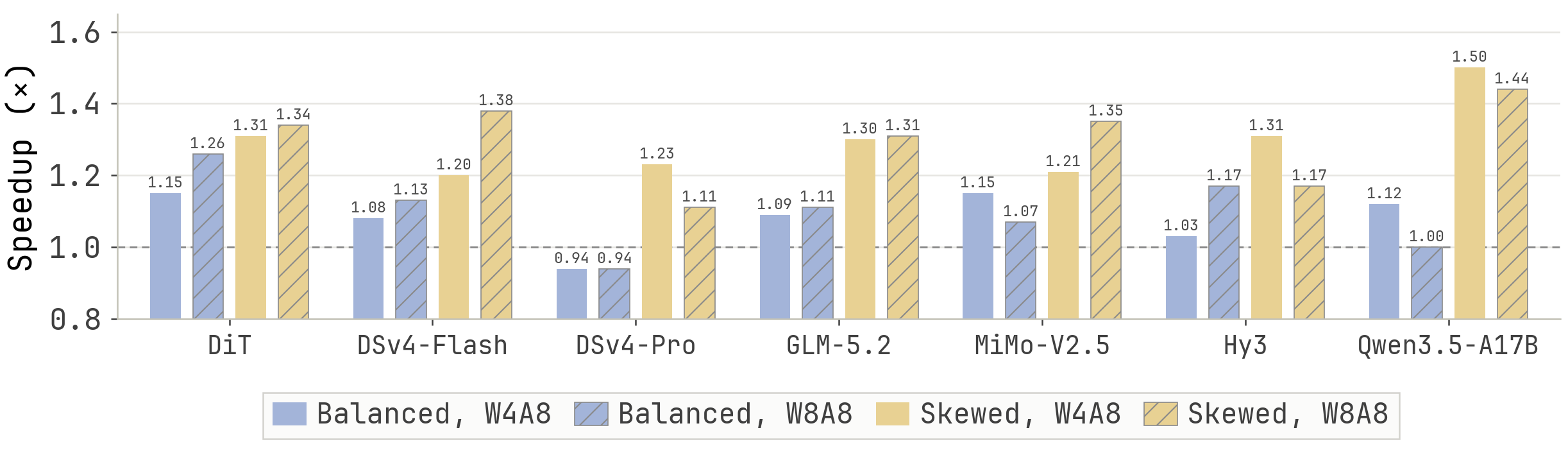}
  \caption{\textbf{MegaMoE kernel speedup} of X-Stage-aware interleaving
over Expert-Wave across models, precisions, and routing distributions.
Each bar shows the mean speedup over three sequence lengths.}
  \Description{Grouped bars showing MegaMoE speedup over Expert-Wave
for each model, precision, and routing distribution, averaged over
three sequence lengths.}
  \label{fig:megamoe-speedup}
\end{figure*}

\textbf{Kernel performance.}
Across all 84 configurations, X-Stage-aware interleaving achieves
a $1.18\times$ geometric-mean, $1.17\times$ median, and $1.62\times$
maximum speedup. Figure~\ref{fig:megamoe-speedup} summarizes these
results by averaging over three sequence lengths.
Skewed routing benefits more because hot experts create longer runs
of aligned Linear-2 epilogues. Several balanced configurations regress
to about $0.94\times$: their communication-side gain is smaller,
while cross-expert interleaving reduces weight-cache locality.
This tradeoff motivates selecting the schedule from both X-Stage
pressure and working-set size rather than always interleaving.

\begin{figure*}[t]
  \centering
  \begin{minipage}[t]{0.48\textwidth}
    \centering
    \includegraphics[width=\linewidth]
    {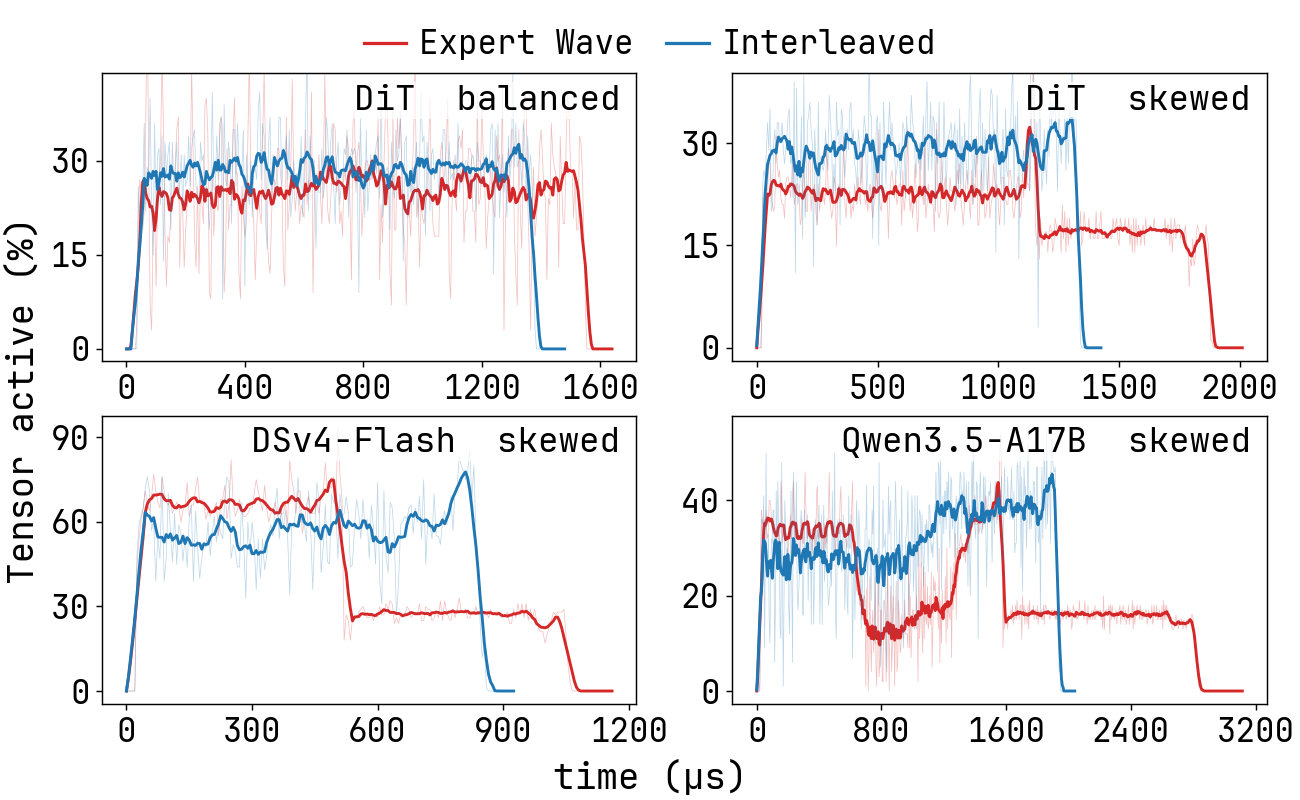}
    \small (a) Tensor-pipe activity.
  \end{minipage}
  \hfill
  \begin{minipage}[t]{0.48\textwidth}
    \centering
    \includegraphics[width=\linewidth]
    {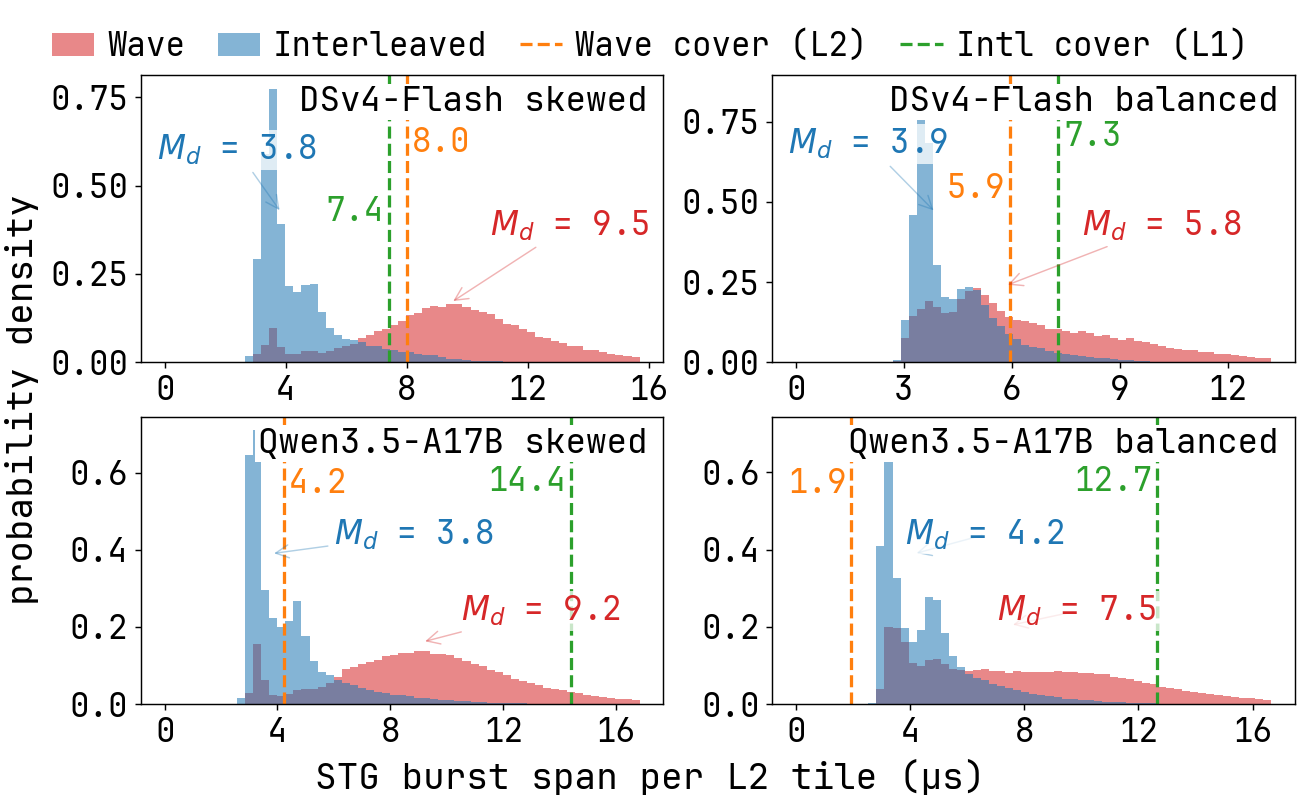}
    \small (b) Per-tile remote-store span.
  \end{minipage}
  \caption{\textbf{MegaMoE mechanism validation.}
  Interleaving shortens low-Tensor-activity intervals and shifts remote-store
  spans from a long, heavy-tailed drain-limited regime toward a narrow
  backpressure-free range.}
  \Description{A Tensor Core activity timeline and distributions of per-tile
  remote-store span before and after MegaMoE interleaving.}
  \label{fig:megamoe-mechanism}
\end{figure*}

\textbf{Headroom and epilogue span.}
If Expert-Wave consumes X-Stage headroom, its aligned spans should approach a
drain-limited scale, while Equation~\ref{eq:megamoe-gap} should move
interleaved spans toward the recovered floor. Figure~\ref{fig:megamoe-mechanism}
shows both effects. For skewed W8A8 routing, the seven Expert-Wave medians range
from $7.9$ to $9.9~\si{\micro\second}$ and exhibit pronounced tails. All seven
interleaved medians narrow to $3.6$--$3.9~\si{\micro\second}$. For three
representative shapes, remote interleaved medians are within
$0.5~\si{\micro\second}$ of matched local-HBM controls. Communication volume is
unchanged, so the shift follows from arrival reshaping rather than less data.

\textbf{Propagation to computation.}
The epilogue span matters only when finite local staging cannot cover it. We
therefore compare a prediction from each tile's measured span and successor
mainloop window against an independent MMA-side wait on
\texttt{tmem\_empty}. Table~\ref{tab:megamoe-stall} reports the result. For six
backpressured Expert-Wave cases, the model preserves the ordering and scale of
the measured $0.65$--$4.31~\si{\micro\second}$ stalls. Interleaving predicts
zero, and measured stalls are at most $0.05~\si{\micro\second}$. DSv4-Pro is a
negative control: its long Linear-2 mainloop already covers the wave span, so
both schedules show negligible MMA stall and little kernel-level benefit.

\begin{table}[t]
  \centering
  \caption{\textbf{Per-tile critical-path test}
  (W8A8, skewed routing, 55{,}808 tokens, 8 GPUs; times in
  \si{\micro\second}). Each entry is prediction/measurement.}
  \label{tab:megamoe-stall}
  \footnotesize
  \setlength{\tabcolsep}{4pt}
  \renewcommand{\arraystretch}{1.1}
  \begin{tabular}{@{}lcc@{}}
    \toprule
    Model & Expert-Wave & Interleaved \\
    \midrule
    DiT-MoE    & 4.63\,/\,4.18 & 0\,/\,0.04 \\
    Qwen3.5    & 4.94\,/\,4.31 & 0\,/\,0.04 \\
    Hy3        & 3.60\,/\,3.08 & 0\,/\,0.05 \\
    MiMo-V2.5  & 1.45\,/\,0.71 & 0\,/\,0.05 \\
    GLM-5.2    & 1.54\,/\,0.92 & 0\,/\,0.04 \\
    DSv4-Flash & 1.47\,/\,0.65 & 0\,/\,0.05 \\
    DSv4-Pro   & 0\,/\,0.06    & 0\,/\,0.04 \\
    \bottomrule
  \end{tabular}
\end{table}

Together, the measurements follow the proposed causal order: interleaving
increases useful drain windows, per-tile issue spans return toward their floor,
MMA-side staging waits disappear, and kernel time improves when the removed
stall was previously critical.

\subsection{UP/Attention Fusion and Supplementary GEMM--A2A Results}
\label{sec:eval-fa}

We fuse the post-attention Ulysses All-to-All with FlashAttention-3 (FA3) and
FlashAttention-4 (FA4), sweeping sequence length $M$. \textbf{FA-only} uses the
same compute path as its fused counterpart; \textbf{Serial} executes that
FlashAttention followed by the same post-attention A2A; and
\textbf{X-Stage-Fused} piggybacks issue on the output-owning role.

We report sender-visible speedup over Serial and the residual

\[
E_{\mathrm{res}}=T_{\mathrm{fused}}-T_{\mathrm{FA}}.
\]

This residual includes issue, address-generation, and measurement effects; it
does not assert that output data is remotely visible at the sender-side timing
boundary.

\begin{table}[t]
  \centering
  \caption{\textbf{Sender-visible FlashAttention--A2A performance.}
  Times are in \si{\micro\second}; speedups use unrounded medians.}
  \label{tab:fa-results}
  \footnotesize
  \setlength{\tabcolsep}{4.2pt}
  \renewcommand{\arraystretch}{1.12}
  \begin{tabular}{@{}llrrrrr@{}}
    \toprule
    Method & Metric / $M$ & 8,192 & 16,384 & 32,768 & 49,152 & 65,536 \\
    \midrule
    \multirow{4}{*}{FA3+A2A}
      & FA-only & 195.2 & 785.5 & 3,236.7 & 7,275.6 & 13,249.4 \\
      & Serial & 295.6 & 968.3 & 3,579.6 & 7,767.9 & 13,936.6 \\
      & \textbf{Fused} & \textbf{207.0} & \textbf{795.5} & \textbf{3,241.5} & \textbf{7,285.7} & \textbf{13,218.3} \\
      & \textbf{Speedup} & \textbf{1.428$\times$} & \textbf{1.217$\times$} & \textbf{1.104$\times$} & \textbf{1.066$\times$} & \textbf{1.054$\times$} \\
    \midrule
    \multirow{5}{*}{FA4+A2A}
      & FA-only & 85.4 & 346.5 & 1,439.4 & 3,281.4 & 5,340.7 \\
      & Serial & 140.9 & 462.4 & 1,660.9 & 3,661.7 & 5,813.0 \\
      & \textbf{Fused} & \textbf{99.4} & \textbf{356.6} & \textbf{1,429.2} & \textbf{3,282.1} & \textbf{5,348.9} \\
      & $E_{\mathrm{res}}$ & $+14.0$ & $+10.1$ & $-10.2$ & $+0.7$ & $+8.2$ \\
      & \textbf{Speedup} & \textbf{1.417$\times$} & \textbf{1.297$\times$} & \textbf{1.162$\times$} & \textbf{1.116$\times$} & \textbf{1.087$\times$} \\
    \bottomrule
  \end{tabular}
\end{table}

\textbf{Sender-visible performance.}
Table~\ref{tab:fa-results} shows maximum speedups of $1.43\times$ for FA3 and
$1.42\times$ for FA4 at $M=8{,}192$. The speedup decreases at longer sequences
because full-attention compute grows faster than the A2A contribution, so even
perfectly hidden sender-side A2A yields
$1+T_{\mathrm{A2A}}/T_{\mathrm{FA}}\rightarrow1$. This is not reduced hiding:
fused time approaches FA-only time.

\begin{figure}[t]
  \centering
  \includegraphics[width=0.98\columnwidth]
  {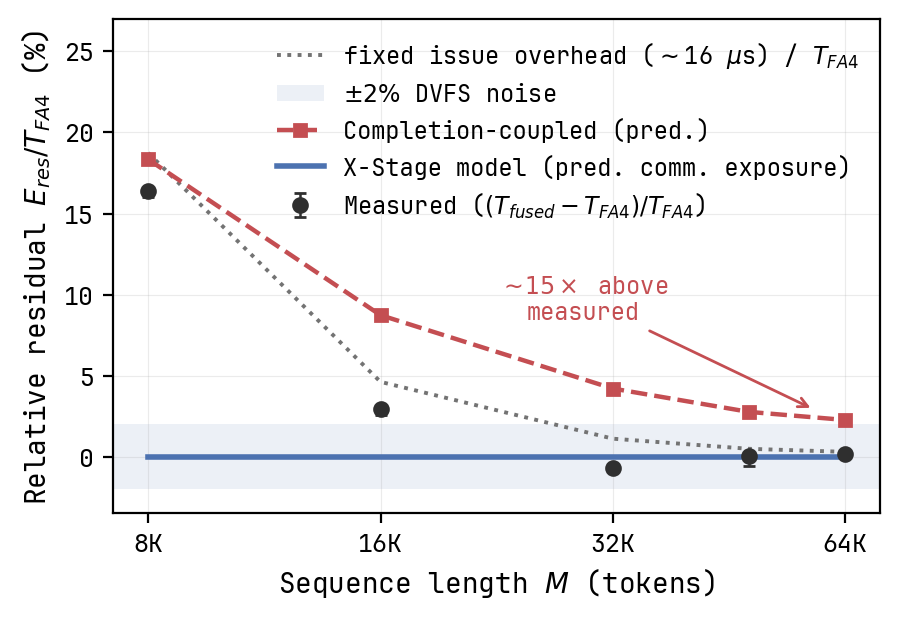}
  \caption{\textbf{Relative sender-visible residual for FA4+A2A.}
  Measurements remain near the issue/noise floor and do not follow the
  completion-coupled growth prediction. The blue band represents the measured
  clock-variation envelope.}
  \Description{Measured relative FlashAttention-4 residual across sequence
  lengths compared with X-Stage and completion-coupled predictions.}
  \label{fig:fa-residual}
\end{figure}

\textbf{Model discrimination.}
Figure~\ref{fig:fa-residual} tests scaling rather than a single point. The
completion-coupled counterfactual predicts an additional residual increasing
from about $16~\si{\micro\second}$ at $M=8{,}192$ to
$120~\si{\micro\second}$ at $M=65{,}536$. FA4 instead remains within
$-10.2$ to $+14.0~\si{\micro\second}$, less than 0.7\% of FA-only time at the
three longest sequences. Slightly negative values arise from subtracting
similar millisecond-scale measurements under clock variation; we do not
interpret them as communication accelerating attention. FA3 exhibits the same
flat-to-slow residual trend. The evidence supports issuer decoupling and rejects
the completion-coupled slope over the evaluated range, while leaving remote
readiness governed by the normal synchronization protocol.

\paragraph{GEMM--All-to-All fusion.}
\label{sec:eval-gemm-a2a}

Ulysses offers another fusion opportunity at the projection--exchange
boundary before attention: the projection GEMM can be fused with the
All-to-All that redistributes its output from sequence to head partitions
\cite{ulysses}. Motivated by this boundary, we evaluate GEMM--All-to-All
fusion on eight GPUs for five sizes from $M=8{,}192$ to $65{,}536$,
with $N=K=4096$. Table~\ref{tab:xstage-gemm-a2a} reports the measured
latencies. Including the final completion step, fusion achieves
$2.14\times$--$2.29\times$ speedup ($2.23\times$ geometric mean) over
the serial path, which includes GEMM+Quant, All-to-All, and the surrounding
permutations. 


\begin{table}[ht]
  \centering
  \caption{\textbf{GEMM--All-to-All fusion on eight GPUs.}
  Latencies in \si{\micro\second}; $N=K=4096$.
  G+Q is the local GEMM+Quant reference. Serial includes the surrounding
  permutations; Fused includes the final completion step.
  Speedup is $T_{\mathrm{Serial}}/T_{\mathrm{Fused}}$.}
  \label{tab:xstage-gemm-a2a}
  \small
  \setlength{\tabcolsep}{2.5pt}
  \renewcommand{\arraystretch}{1.12}
  \begin{tabular*}{\columnwidth}{@{\extracolsep{\fill}}rrrrr@{}}
    \toprule
    $M$ & G+Q & Serial & Fused & Speedup \\
    \midrule
     8,192 &   209.1 &   529.5 &   246.9 & $2.145\times$ \\
    16,384 &   413.8 & 1,033.2 &   464.1 & $2.226\times$ \\
    32,768 &   831.2 & 2,048.0 &   894.0 & $2.291\times$ \\
    49,152 & 1,227.7 & 3,076.4 & 1,366.6 & $2.251\times$ \\
    65,536 & 1,643.9 & 4,119.3 & 1,823.8 & $2.259\times$ \\
    \bottomrule
  \end{tabular*}
\end{table}

Beyond compute--communication overlap, fusion eliminates the separate
permutation preceding NCCL All-to-All. Removing this additional work
allows the overall speedup to exceed the $2\times$ ceiling achievable
by overlapping two stages with unchanged costs.

\subsection{TP/GEMM--AllReduce: Fusion Performance}
\label{sec:eval-tp}

We compare the fused GEMM--AllReduce operation with serial GEMM+Quant
followed by AllReduce on eight GPUs. Table~\ref{tab:xstage-gemm-ar} reports five
matrix sizes with $N=K=4096$. The local GEMM+Quant measurement provides the
producer reference; speedup is the serial operation time divided by the fused
operation time.

\begin{table}[ht]
  \centering
  \caption{\textbf{GEMM--AllReduce fusion on eight GPUs.}
  Latencies in \si{\micro\second}; $N=K=4096$. G+Q denotes GEMM+Quant.
  Each latency is the median of three runs.}
  \label{tab:xstage-gemm-ar}
  \small
  \setlength{\tabcolsep}{2.5pt}
  \renewcommand{\arraystretch}{1.12}
  
  \begin{tabular*}{\columnwidth}{@{\extracolsep{\fill}}rrrrr@{}}
    \toprule
    $M$ & G+Q & Serial & Fused & Speedup \\
    \midrule
     8,192 &   227.4 &   567.5 &   357.3 & $1.588\times$ \\
    16,384 &   437.2 & 1,032.3 &   615.7 & $1.677\times$ \\
    32,768 &   860.1 & 1,975.3 & 1,130.0 & $1.748\times$ \\
    49,152 & 1,266.2 & 2,921.1 & 1,798.9 & $1.624\times$ \\
    65,536 & 1,690.7 & 3,920.3 & 2,360.2 & $1.661\times$ \\
    \bottomrule
  \end{tabular*}
\end{table}

Fusion achieves $1.59\times$--$1.75\times$ speedup, with a
$1.66\times$ geometric mean across these five configurations. The fused
operation remains longer than the local producer in every case, so the result
does not imply that all communication is hidden behind GEMM. We report this
as a TP fusion performance result; the timing comparison alone does not
quantify X-Stage occupancy or isolate all sources of the gain.

\section{Discussion and Extensions}
\label{sec:discussion}

\textbf{Platform portability.}
X-Stage abstracts the effective post-issue behavior of the GPU
communication path. Its parameters
$(T_{\mathrm{iss}}^{0},\mathcal{R},\mathcal{Q})$ depend on the platform,
remote-store path, and producer configuration. Transferring the model
requires validating these parameters under the new configuration.
When producer alignment varies within a modeled period, a finer arrival
trace may be needed to capture short-lived backlog peaks.

\textbf{Extension to other fusion patterns.}
The same formulation offers a way to analyze other fused kernels that
interleave device-initiated remote stores with useful computation.
Potential examples include push-based All-Gather and Reduce-Scatter
phases coupled with independent tile computation. Each implementation
would supply its per-phase burst volumes, producer alignment, and
compute intervals, while retaining the collective's reduction and
data-readiness dependencies. The model could then identify where
arrivals accumulate pressure and where existing computation provides
a recovery window. This extends the analysis through the execution
pattern rather than through a particular operator or parallelism label.

\textbf{Completion semantics.}
X-Stage predicts sender-side issue recovery and backpressure.
Local-buffer reuse and remote consumption remain governed by the
implementation's completion and synchronization protocols. For fusion
patterns in which communication supplies the next computation, the
sender-side prediction must also be combined with receiver readiness
to determine the compute critical path.

\section{Related Work}
\label{sec:related-work}

\textbf{Device-initiated communication and fusion.}
Peer mappings and symmetric memory let kernels embed one-sided communication
inside compute pipelines \cite{nvshmem,nvshmem_demystify}. CoCoNet, FLUX,
Comet, GC3, TileLink, MegaScale-MoE, and ParallelKittens expose fine-grained
fusion through task decomposition, scheduling, specialized roles, and
device-side synchronization
\cite{coconet,flux,comet,gc3,tilelink,megascale,parallelkittens}.
These systems establish issue roles and remote readiness. X-Stage is
complementary: it models the finite sender-visible state after issue and uses
that state to choose between arrival reshaping and issuer reuse.

\textbf{MoE communication.}
MoE systems reduce expert-parallel communication with grouped GEMM, sparse
computation, placement, load balancing, and overlap
\cite{MegaBlocks,fastermoe,eplb,metro,realb,lancet,fsmoe}.
DeepGEMM MegaMoE and UniEP fuse \moedispatch{}, expert computation, and
\moecombine{} into persistent kernels \cite{deepgemm,uniep}. Our MegaMoE
transformation does not change routing, placement, dependencies, or bytes; it
redistributes ready Linear-1 work to control when an unchanged \moecombine{}
volume enters X-Stage.

\textbf{Tensor-parallel fusion.}
FLUX, Comet, TileLink, and related fused collective kernels overlap GEMM tiles
with TP communication through specialized roles, chunking, and device-side
coordination \cite{flux,comet,tilelink}. X-Stage does not prescribe one TP
chunk size or role assignment. It provides the missing sender-side audit for
how those choices change aligned burst volume, useful gap, backlog, and the
capacity guard.

\textbf{Sequence-parallel attention and performance models.}
Ulysses, Ring Attention, and USP distribute long-sequence attention using
All-to-All or ring communication \cite{ulysses,ring_attention,usp}; the
FlashAttention family provides IO-aware tiled compute pipelines
\cite{fa,fa2,fa3,fa4}. We fuse the post-attention Ulysses All-to-All and use the
next Q-loop as a measured post-issue drain window. Classical models such as
Roofline, LogP/LogGP, and Little's law describe aggregate throughput, messaging
costs, or average occupancy \cite{roofline,logp,loggp,little1961}. The
Burst--Gap model instead targets sender-visible issue recovery and finite
post-issue capacity inside a fused kernel.

\section{Conclusion}
\label{sec:conclusion}

X-Stage provides an architectural abstraction of finite post-issue
decoupling in GPU remote-store execution. We characterize how burst
arrivals consume headroom and how that headroom recovers during useful
computation. Our calibrated Burst--Gap model translates these architectural
properties into quantitative guidance for GPU kernel programming,
identifying when to reshape arrivals and when to exploit existing compute
windows. Evaluation across representative EP, TP, and UP workloads
demonstrates fusion benefits, while targeted measurements validate the
predicted recovery and backpressure mechanisms. X-Stage makes finite
post-issue headroom an explicit consideration in the design of
communication--computation overlap.

\newpage

\FloatBarrier
\bibliographystyle{ACM-Reference-Format}

\bibliography{references}

\clearpage
\appendix

\section{MegaMoE Scheduler and Local-Staging Mapping}
\label{app:megamoe}

\subsection{Interleaving and Average-Rate Feasibility}
\label{app:megamoe-rate}
In the drain-limited branch of Equation~\ref{eq:burst-gap}, the burst period is
fixed by $\mathcal{V}/\mathcal{R}$; scheduling determines how much of that
period is useful computation and how much is sender-visible issue stall.
Substituting Equation~\ref{eq:megamoe-gap} into the Burst--Gap relation yields

\begin{equation}
\Delta T_{\mathrm{iss}}^{\mathrm{int}}
=
\left[
\Delta T_{\mathrm{iss}}^{\mathrm{wave}}
-\frac{n_1}{n_2}T_{\mathrm{L1}}^{\mathrm{mma}}
\right]_{+}.
\label{eq:megamoe-overhead}
\end{equation}

Interleaving therefore exchanges one unit of previously grouped Linear-1 work
for approximately one unit of reduced issue stall until the recovered floor is
reached. This exchange is enabled by post-issue
progress: useful work inserted between issue phases can replace sender-visible
stall without changing communication volume. A necessary average-rate feasibility condition for complete
modeled recovery is

\begin{equation}
\frac{n_2\mathcal{V}}
{n_2(T_{\mathrm{iss}}^{0}+T_{\mathrm{L2}}^{\mathrm{mma}})
 +n_1T_{\mathrm{L1}}^{\mathrm{mma}}}
\le\mathcal{R}.
\label{eq:megamoe-sufficient}
\end{equation}

This condition depends on the calibrated X-Stage rate, aggregate burst, tile
mix, and mainloop times, but not on the scheduler's dependency lead.
If it fails, reordering alone
cannot fully restore headroom. If it holds, the concrete tile order must still
pass the prefix-backlog recurrence and capacity guard; the dependency lead can
change those prefixes even when the average-rate condition is satisfied.

\subsection{Scheduler and Local-Staging Mapping}

\begin{algorithm}[ht]
  \DontPrintSemicolon
  \SetAlgoNoLine
  \SetAlgoNoEnd
  \footnotesize
  \caption{Cross-wave Linear-1/Linear-2 interleaving.}
  \label{alg:megamoe-interleave}
  \Proc{\textnormal{\textsc{Scheduler}}$(g,r)$}{
    $k_1\leftarrow g$;\quad $k_2\leftarrow g$\;
    \While{$k_1<P P_1$ \textbf{or} $k_2<P P_2$}{
      $p_1\leftarrow\lfloor k_1/P_1\rfloor$
        \quad ($+\infty$ if Linear-1 is done)\;
      $p_2\leftarrow\lfloor k_2/P_2\rfloor$\;
      \eIf{$k_2<P P_2$ \textbf{and} $p_2+D\le p_1$}{
        \textbf{emit} Linear-2 tile $(p_2,k_2,r)$\;
        $k_2\leftarrow k_2+G$\;
      }{
        \textbf{emit} Linear-1 tile $(p_1,k_1,r)$\;
        $k_1\leftarrow k_1+G$\;
      }
    }
  }
\end{algorithm}

Here $g$ and $G$ identify a strided scheduler instance, $r$ is the role in a
paired CTA, $P_i$ is the number of tile pairs per row, and $D$ is the minimum
row lead of Linear-1 over Linear-2. The implementation converts scheduler rows
to expert-local M/N tile coordinates and checks \moedispatch{} and ring-slot
readiness before emission.

The dependency-derived anti-starvation knee is approximately

\[
D_{\mathrm{knee}}
=
\left\lceil
\frac{b_{\mathrm{n}}\mathcal{K}_{\mathrm{SM}}}
{2h_{\mathrm{inter}}}
\right\rceil .
\]

This threshold comes from cross-SM Linear-1 readiness, not X-Stage draining.
Increasing $D$ beyond the knee does not change the long-run tile ratio or the
rate condition; it only increases the Linear-2 tail and live expert-weight
working set.

For the mechanism test, let $T_{\mathrm{RS}}$ be one tile's measured
sender-visible remote-store span and let
$T_{\mathrm{cover}}^{\mathrm{mma}}$ be the mainloop time before its local
staging slot must be reused. The predicted MMA-side stall is

\[
\Delta t
=
\left[
T_{\mathrm{RS}}-T_{\mathrm{cover}}^{\mathrm{mma}}
\right]_{+}.
\]

The successor is generally a Linear-2 mainloop in Expert-Wave and may be a
longer Linear-1 mainloop after interleaving. Table~\ref{tab:megamoe-stall}
compares this prediction with independent \texttt{tmem\_empty} instrumentation.

\section{Ulysses Dataflow and FlashAttention Traversal}
\label{app:fa-context}

\begin{figure*}[t]
  \centering
  \includegraphics[width=\textwidth]{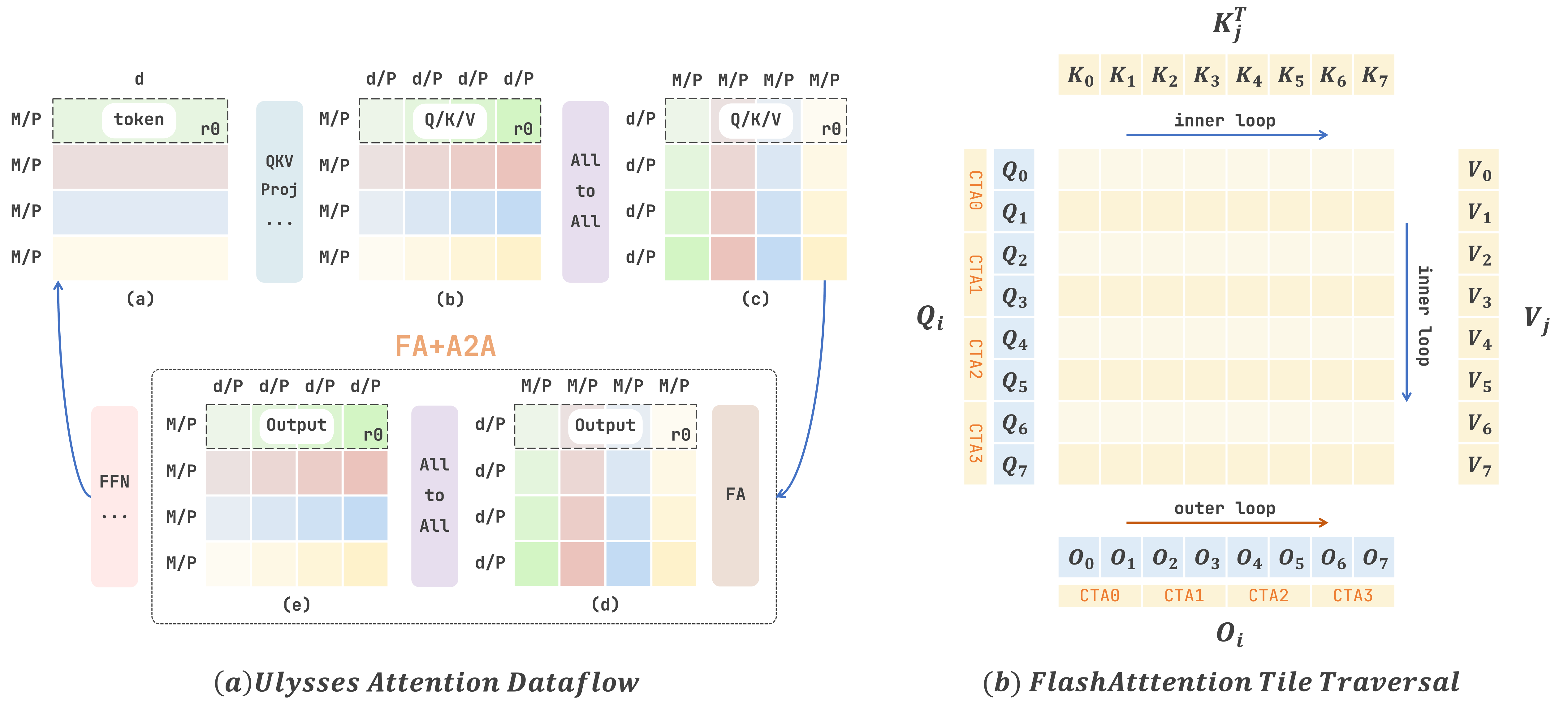}
  \caption{\textbf{Where the natural drain window comes from.}
  (a) Ulysses uses a pre-attention All-to-All to form head partitions and a
  post-attention All-to-All to restore sequence partitions; the FlashAttention
  fusion targets the latter. (b) A FlashAttention CTA completes a full traversal of K/V tiles
  before producing its next output tile, placing a complete Q-loop between
  consecutive output-side remote-store bursts.}
  \Description{Ulysses data-layout transformations alongside the
  FlashAttention Q-loop and output-tile structure.}
  \label{fig:fa-context}
\end{figure*}

Figure~\ref{fig:fa-context} connects Ulysses data redistribution
with FlashAttention's tile traversal. The pre-attention All-to-All
converts sequence partitions into head partitions, allowing each
rank to compute attention over the full sequence for its assigned
heads. The post-attention exchange restores sequence partitions.

Our FlashAttention fusion attaches remote stores to the output
boundary. When a CTA processes successive query tiles, the next
tile's Q-loop traverses the K/V tiles before producing another
output burst. At a fixed output-tile shape, increasing sequence
length extends this compute interval without increasing the bytes
per output burst, providing more time for X-Stage to drain.

\section{Conservative FlashAttention Recovery Bound}
\label{app:fa-bound}

\subsection{Capacity and Recovery Checks for FA4}
\label{app:fa-calibration}

The model audit explains why piggybacking is sufficient in our evaluated
region. For FlashAttention-4,
$b_{\mathrm{m}}=128$, head dimension $d=128$, and bf16 output give 32~KiB per
producer. Even if all $\mathcal{K}=148$ producers align, the aggregate burst is
about 4.6~MiB and
$b^{\mathrm{iso}}\approx4.1$~MiB
$<\mathcal{Q}\approx4.25$~MiB. Under these point estimates, a recovered burst
fits within the calibrated effective capacity. Its worst-case recovery gap is about
$\mathcal{G}^{*}=6.0~\si{\micro\second}$, whereas the measured Q-loop gap is
about $80~\si{\micro\second}$ even at the smallest evaluated sequence length
$M=8{,}192$. Thus $\rho_n<1$, $\Delta b_n<0$, and the recurrence restores
$q_{n+1}=0$: the measured downstream pressure can dissipate during the
Q-loop, so headroom recovers before the next output boundary.

\subsection{Sequence-Length Recovery Bound}

A FlashAttention output tile performs both $QK^{T}$ and $PV$ over all K/V
tiles. For tile shape $b_{\mathrm{m}}\times b_{\mathrm{n}}$, head dimension
$d$, and sequence length $M$, the work is approximately

\[
C_{\mathrm{FA}}
=
4b_{\mathrm{m}}b_{\mathrm{n}}d
\left\lceil\frac{M}{b_{\mathrm{n}}}\right\rceil.
\]

The measured Q-loop gap has a positive fixed component and an approximately
linear slope. Dropping the positive component gives

\[
\mathcal{G}_{\mathrm{Qloop}}(M)
\ge
\sigma\left\lceil\frac{M}{b_{\mathrm{n}}}\right\rceil,
\qquad
\sigma\approx0.79~\si{\micro\second}\ \text{per K/V tile}.
\]

A sufficient tile-aligned threshold for recovery is therefore

\[
M_{\mathrm{ub}}^{*}
=
b_{\mathrm{n}}
\left\lceil\frac{\mathcal{G}^{*}}{\sigma}\right\rceil.
\]

For the worst aligned case $\mathcal{K}=148$, the calibrated
$\mathcal{G}^{*}\approx6.0~\si{\micro\second}$ gives
$M_{\mathrm{ub}}^{*}\approx1.0$K tokens. The true threshold can be smaller
because the omitted fixed gap is positive, fewer producers may align, and the
compute critical path can tolerate residual epilogue delay covered by local
staging.

\section{Additional Measurement Details}
\label{app:methodology}

For each MegaMoE cell, an untimed invocation performs JIT compilation and
warm-up. We collect 20 barrier-aligned CUDA-event trials per rank, take the
maximum latency across ranks for each trial, and report the median. Scheduler
order is rotated across cells, and anomalous cells are repeated in paired,
order-rotated runs.

FA4 uses CUDA Graph replay because its CuteDSL Python launch overhead is
comparable to short-sequence device time. FA3 uses steady-state CUDA-event
timing. Long-sequence residuals subtract similar millisecond-scale
measurements, so small absolute clock changes are amplified. We monitor SM
clocks, rotate execution order, repeat affected points in clock-stabilized
runs, and treat small negative residuals as noise. Back-to-back replay tests
sustained sender-side behavior; measurement blocks are separated using
\texttt{torch.distributed.barrier()} with the NCCL backend and
\texttt{torch.cuda.synchronize()}, outside the CUDA-event timing
intervals. Output correctness checks separately invoke
\texttt{hdl.barrier(channel=0)}, where \texttt{hdl} is the PyTorch
symmetric-memory handle, before reading the received output.

\end{document}